# Unconditionally Secure Quantum Bit Commitment Protocols Based on Correlation Immune Boolean Function

Li Yang[*] and Bao Li

State Key Laboratory of Information Security, Graduate University of Chinese Academy of Sciences, Beijing 100049, China

A class of quantum protocols of bit commitment is constructed based on the nonorthogonal states coding and the correlation immunity of some Boolean functions. The binding condition of these protocols is guaranteed mainly by the law of causality and the concealing condition is guaranteed by the indistinguishability between nonorthogonal quantum states and the correlation immunity of Boolean functions. We also give out an oblivious transfer protocol based on two-nonorthogonal states coding and build a bit commitment protocol on top of it. The relationship between these protocols and the well known no-go theorem is also discussed in details.

Bit commitment is a basic building block of many cryptography protocols [1], especially those related with two-party secure computations [2]. Unfortunately, it can be proved that there is no classical bit commitment protocol which satisfies unconditionally secure conditions in both binding and concealing. It had been believed that with the help of quantum mechanics one can construct unconditionally secure bit commitment protocols. Unfortunately again, D. Mayers [3] and H. –K. Lo and H. F. Chau [4] proved their well known no-go theorem which declares that unconditionally secure bit commitment protocols based on quantum mechanics are all ruled out. Since then researchers have presented various compromise protocols [5,6]

Generally speaking, bit commitment protocol is a function $f : \{0,1\} \times X \to Y$, where $X$ and $Y$ are two finite set. An encryption of $b \in \{0,1\}$ is a value of $\{f(b,x), x \in X\}$. A bit commitment protocol must satisfy two conditions as follows: (1) Concealing. The receiver cannot get $b$ from $f(b,x)$. (2) Binding. The committer can open $f(b,x)$ via sending $x$ to the receiver, but he cannot open it both as 0 and as 1. Our basic idea of constructing secure bit commitment protocols is to combine nonorthogonal states coding and correlation immune Boolean function [1] together to construct unconditionally secure quantum encryption functions, and then, based on those functions, we construct unconditionally secure quantum bit commitment protocols.

Consider a committer Alice and a receiver Bob communicate over a quantum channel as well as a classical channel. We limit our attention to two class states coding: two nonorthogonal states coding, or Bennett 1992[7] (B92)-like coding, and four states coding, or Bennett-Brassard 1984 [8] (BB84)-like coding. The key technique of our protocols is the adoption of correlation immune Boolean function which is defined as follows [1]: Let $A_1, A_2, \ldots, A_n$ be independent binary variables, each taking on the values 0 or 1 with probability $\frac{1}{2}$. A Boolean function $F(a_1, a_2, \ldots, a_n)$ is $n_0^{th}$-order correlation immune if for each subset of $n_0$ random variables $A_{i_1}, A_{i_2}, \ldots, A_{i_{n_0}}$ with $1 \leq i_1 < i_2 < \cdots < i_{n_0} \leq n$, the random variable $B = F(A_1, A_2, \ldots, A_n)$ is statistically independent of the random vector $(A_{i_1}, A_{i_2}, \ldots, A_{i_{n_0}})$.

It can be seen that in the proof of the no-go theorem the authors have not pay attention to the case in which though the deference between $\rho_0^B$ and $\rho_1^B$ is infinitesimal but they still can be distinguished. We shall give out a detailed explanation of this problem later in '*Discussions*'.

*Quantum Encryption Function*. Here we discuss quantum encryption function related with bit commitment. For any vector $a \in \{0,1\}^k$, we can choose $\{|\Psi_0\rangle, |\Psi_1\rangle\}$, where $0 < \langle \Psi_0 | \Psi_1 \rangle < 1$, to construct a weak one-way function [9] which maps a classical variable to a qubit sequence [10]:

$$f : a = (a_1, a_2, \ldots, a_k) \rightarrow |\Psi_{a_1}\rangle, |\Psi_{a_2}\rangle, \ldots, |\Psi_{a_k}\rangle. \tag{1}$$

When $k$ is sufficiently large, it becomes a strong one-way function [9]. This construction is rather simple and direct, but it cannot be used directly to cryptology because the one-way function used in cryptology must be an verifiable function. One verifiable function has been studied more than twenty years ago as the first quantum bit commitment protocol [8]:

$$\begin{aligned} f : a &= (a_1, a_2, \ldots, a_k) \\ &\rightarrow |\Psi_{a_1^{(1)}}^{(a_1)}\rangle, \ldots, |\Psi_{a_n^{(1)}}^{(a_1)}\rangle, |\Psi_{a_1^{(2)}}^{(a_2)}\rangle, \ldots, |\Psi_{a_n^{(2)}}^{(a_2)}\rangle, \ldots, |\Psi_{a_1^{(k)}}^{(a_k)}\rangle, \ldots, |\Psi_{a_n^{(k)}}^{(a_k)}\rangle. \end{aligned} \tag{2}$$

Where

$$a^{(i)}\Big|_{i=1,2,\ldots,k} = (a_1^{(i)}, a_2^{(i)}, \ldots, a_n^{(i)})\Big|_{i=1,2,\ldots,k} \in \{0,1\}^n \tag{3}$$

is picked randomly for each computation, and

$$\{|\Psi_0^{(0)}\rangle, |\Psi_1^{(0)}\rangle, |\Psi_0^{(1)}\rangle, |\Psi_1^{(1)}\rangle\} = \{|0\rangle, |1\rangle, |+\rangle, |-\rangle\}. \tag{4}$$

This quantum function is easy to be verified if one has $a$ and $\{a^{(i)} | i = 1, 2, \ldots, k\}$. It can be see that this quantum function is an 'one-way' function with a key, that is so called encryption function. We can find that the number of output qubits for every input bit increases quickly. We guess that the number of output qubits corresponding to 1 bit input for a verifiable quantum encryption function with verification security $O(2^{-n})$ is $O(n)$, the number $n$ is so called security parameter.

It is well known that this encryption function can be opened as different values; this leads to the first bit commitment protocol insecure. There is an altered function as follows:

$$f : a = (a_1, a_2, \ldots, a_k) \\
\to \left|\Psi_{a_1}^{(a_1^{(1)})}\right\rangle, \ldots, \left|\Psi_{a_1}^{(a_n^{(1)})}\right\rangle, \left|\Psi_{a_2}^{(a_1^{(2)})}\right\rangle, \ldots, \left|\Psi_{a_2}^{(a_n^{(2)})}\right\rangle, \ldots, \left|\Psi_{a_k}^{(a_1^{(k)})}\right\rangle, \ldots, \left|\Psi_{a_k}^{(a_n^{(k)})}\right\rangle. \tag{5}$$

It cannot be opened as different input value. Unfortunately, one can get the value of $a$ via Breidbart attack before he received $\{a^{(i)} | i = 1, \cdots, k\}$.

Now let us consider another essential property of quantum encryption function: inalterability. Because of the character of quantum states, especially the existence of attacks based on entanglement, we have to inspect carefully the inalterability of a quantum function designed for cryptographic purpose. It is obvious that the existence of verifiable and unalterable encryption function means directly bit commitment.

We now present this kind of encryption functions as follows:

1. B92-like scheme. Choose $\{|\Psi_0\rangle, |\Psi_1\rangle | 0 < \langle \Psi_0 | \Psi_1 \rangle < 1\}$ and $n_0^{th}$-order correlation immune Boolean functions F. For any input $b \in \{0,1\}$, picks randomly $a^{(i)}\big|_{i=1,2,\ldots,m} \in \{0,1\}^n$ satisfy $F(a^{(i)})\big|_{i=1,2,\ldots,m} = b$. The encryption function is

$$\text{Blob}_2 : b \to \left|\Psi_{a_1^{(1)}}\right\rangle, \ldots, \left|\Psi_{a_n^{(1)}}\right\rangle, \ldots, \left|\Psi_{a_1^{(m)}}\right\rangle, \ldots, \left|\Psi_{a_n^{(m)}}\right\rangle. \tag{6}$$

2. BB84-like scheme. Choose $\{|0\rangle, |1\rangle, |+\rangle, |-\rangle\}$, and $n_0^{th}$-order correlation immune Boolean functions F. For any input $b \in \{0,1\}$, picks randomly $a^{(i)}\big|_{i=1,2,\ldots,m} \in \{0,1\}^n$ satisfy $F(a^{(i)})\big|_{i=1,2,\ldots,m} = b$ and choose $\{\tilde{a}^{(i)} \in \{0,1\}^n | i = 1, 2, \ldots, m\}$ randomly and independently. The encryption function is

$$\text{Blob}_2 : b \to \left|\Psi_{a_1^{(1)}}^{\tilde{a}_1^{(1)}}\right\rangle, \ldots, \left|\Psi_{a_n^{(1)}}^{\tilde{a}_n^{(1)}}\right\rangle, \ldots, \left|\Psi_{a_1^{(m)}}^{\tilde{a}_1^{(m)}}\right\rangle, \ldots, \left|\Psi_{a_n^{(m)}}^{\tilde{a}_n^{(m)}}\right\rangle. \tag{7}$$

Where

$$\{|\Psi_0^{(0)}\rangle, |\Psi_1^{(0)}\rangle, |\Psi_0^{(1)}\rangle, |\Psi_1^{(1)}\rangle\} = \{|0\rangle, |1\rangle, |+\rangle, |-\rangle\}. \tag{8}$$

It can be seen that $\text{Blob}_2$ and $\text{Blob}_4$ are weak encryption functions; 'weak' means the failure probability of guessing the value $b$ does not approaches 0 exponentially. The valuable properties of them are verifiability and inalterability. Let us delay the proof of its verifiability and inalterability to the security analysis of corresponding bit commitment protocols. It is clear that we can construct two strong encryption functions:

$$A(b_1, b_2, \ldots, b_n) = \text{Blob}_2(b_1) \text{Blob}_2(b_2) \cdots \text{Blob}_2(b_n), \tag{9}$$

$$\mathrm{B}(b_1, b_2, \ldots, b_n) = \mathrm{Blob}_4(b_1)\, \mathrm{Blob}_4(b_2) \cdots \mathrm{Blob}_4(b_n), \tag{10}$$

provided $n$ is sufficiently large. Where 'strong' means the probability of successful guess of value $b$ approaches 0 exponentially.

*Bit Commitment Protocol: B92-Like Case.* Committer Alice and receiver Bob choose

$$\left\{ |\Psi_0\rangle, |\Psi_1\rangle \,\Big|\, \frac{1}{2} + \delta \leq |\langle \Psi_0 | \Psi_1 \rangle|^2 \leq 1 - \delta; 0 < \delta \leq \frac{1}{4} \right\} \tag{11}$$

Let $|\langle \Psi_0 | \Psi_1 \rangle| = \cos A$. It is known that the success probability of distinction between these two states is $p_A = 1 - \cos A$ [11,12], then $1 - \sqrt{1-\delta} \leq p_A \leq 1 - \sqrt{\frac{1}{2} + \delta}$. No further improvement is possible [12]. The protocol is as follows.

Commit phase:

1. Alice and Bob choose an $n_0^{th}$-order correlation immune Boolean functions F, where $n_0$ satisfies $n - n_0 \sim \mathrm{O}(1)\ (n \to \infty)$.

2. Alice picks randomly $a^{(i)}\big|_{i=1,2,\ldots,m} \in \{0,1\}^n$, which satisfy $\mathrm{F}(a^{(i)})\big|_{i=1,2,\ldots,m} = b$. Where $b \in \{0,1\}$ is the value Alice committed.

3. Alice generates $\mathrm{Blob}_2(b)$ (see formula (6)) as an encryption function (or, blob) of her commitment and sends it to Bob.

Open phase:

4. Alice unveil $\{a^{(i)} \in \{0,1\}^n \,|\, i = 1, 2, \ldots, m\}$ to Bob.

5. Bob verifies that the blob he has received is really coded by $\{a^{(i)} \,|\, i = 1, 2, \ldots, m\}$, and then calculates $\left(b^{(i)} = \mathrm{F}(a^{(i)})\right)\big|_{i=1,2,\ldots,m}$. If $b^{(i)}\big|_{i=1,2,\ldots,m} = b$, he accepts $b$ as the value Alice committed.

Security Analysis: Binding Property. This corresponds to the verifiability and inalterability of encryption function $\mathrm{Blob}_2(b)$. We consider first verifiability, that means we can find $b \neq b'$ with a probability approaches 1 exponentially. It can be seen that for any $a'^{(i)}$ and $a^{(j)}$ which satisfy $\mathrm{F}(a^{(i)}) = b$ and $\mathrm{F}(a'^{(j)}) = b'$ we have $\mathrm{W}_H\left(a'^{(i)} \oplus a^{(j)}\right) \geq 1$. Where $\mathrm{W}_H(a)$ is the Hamming weight of a. If Alice unveil $\left(b', a'^{(i)}\big|_{i=1,2,\cdots,m}\right)$, $\mathrm{F}(a'^{(i)})\big|_{i=1,2,\ldots,m} = b' \neq b$, then $W_H(a'^{(i)} \oplus a^{(i)}) \geq 1$. After performing local operations to $\text{Blob}_2(b)$, Bob can find $b' \neq b$ with probability $P_B \geq 1 - (\sin^2 A)^m$. $P_B \to 1$ (Exponentially) means that for any positive real parameter $\alpha$, we can choose a parameter $m$ which satisfies $P_B > 1 - e^{-\alpha}$. This means $m > -\dfrac{\alpha}{2\ln(\sin A)} \propto O(\alpha)$. we provide here Alice cannot alter the blob remotely.

Let us consider next the inalterability of $\text{Blob}_2(b)$. It can be seen that when Alice changes the value of $b$ she has to change at least one qubit between $|\Psi_0\rangle$ and $|\Psi_1\rangle$ in each $n$-qubit string. If Alice can realize this kind of change via remote operation, she can construct a superluminal signaling scheme with a success probability larger than $1 - \cos^{2m} A$. Based on the law of causality we conclude that the protocol is secure under all kinds of remote (EPR) attack related with quantum entanglement.

An attack strategy which does not violate the law of causality is: Alice measures probe qubits in her hands, which leads the state of 'blob-probe' system collapses to a definite blob state and a definite probe state. Based on the law of causality we know that Alice cannot control the final state of blob, though she can know it remotely. It is obvious that in this case there are 50% $n$-qubit strings are in the states corresponding to committing 0, and the others in the states committing 1. Suppose Alice is so lucky that after the measurement there is only 1 bit for each wrong $n$-qubit string has to be changed, the probability of Alice's failure is

$$\sum_{k=0}^{m} C_m^k \left(\frac{1}{2}\right)^m \left[1 - (\cos^2 A)^{m-k}\right] = 1 - \left(\frac{1 + \cos^2 A}{2}\right)^m \to 1 \tag{12}$$

To state succinctly, the protocol described is an unconditionally secure binding protocol.

Security Analysis: Concealing Property. Correlation immune Boolean function has only two candidates for $n_0 = n - 1$ case: $F(a) = a_1 \oplus a_2 \oplus \cdots \oplus a_n \oplus c$, where $c \in \{0, 1\}$ is a const. Consider of the menace of weight attack, we know this kind of Boolean function will affects the security of the protocols, so we study only $n_0 < n - 1$ case. In this case, Bob will not be successful except he has get more than $n_0$ components of a vector $a \in \{a^{(i)} | i = 1, \cdots, m\}$. It can be seen that for each $n$-qubit string Bob's failure probability is $p_A^{(n_0)} = \sum_{k=0}^{n_0} C_n^k p_A^k (1 - p_A)^{n-k}$, then Bob's successful cheat probability is $P_A = 1 - \left(p_A^{(n_0)}\right)^m$. The concealing condition $P_A \to 0$ (exponentially) means $p_A^{(n_0)} > \sqrt[m]{1 - e^{-\beta}}$. According to De Moivre-Laplace theorem,

for large $n$, we have

$$p_A^{(n_0)} = \frac{1}{\sqrt{2\pi}} \int_{-\lambda_1 \sqrt{n}}^{\lambda_2 \sqrt{n}} e^{-\frac{t^2}{2}} dt, \tag{13}$$

where $\lambda = \frac{n_0}{np_A}$, $\lambda_1 = \sqrt{\frac{p_A}{1-p_A}}$, $\lambda_2 = (\lambda-1)\lambda_1$. Then we can get

$$p_A^{(n_0)} = \frac{1}{2}\left[\text{erf}(\lambda_1 \sqrt{\frac{n}{2}}) + \text{erf}(\lambda_2 \sqrt{\frac{n}{2}})\right], \tag{14}$$

where $\text{erf}(z)$ is the error function. Neglecting higher-order terms of the error function $\text{erf}(z)$, we have

$$p_A^{(n_0)} = \frac{1}{\sqrt{2\pi n}}\left[\frac{1}{\lambda_1} e^{-\frac{1}{2}\lambda_1^2 n} + \frac{1}{\lambda_2} e^{-\frac{1}{2}\lambda_2^2 n}\right]. \tag{15}$$

It can be seen that $p_A^{(n_0)} > \sqrt[m]{1-e^{-\beta}}$ will holds if and only if $n > O(\beta)$ as $\beta$ is sufficiently large.

*Bit Commitment Protocol: BB84-Like Case.*
  Commit Phase:

1. Alice and Bob choose $\{|0\rangle, |1\rangle, |+\rangle, |-\rangle\}$, where $|+\rangle = \frac{1}{\sqrt{2}}(|0\rangle + |1\rangle)$, $|-\rangle = \frac{1}{\sqrt{2}}(|0\rangle - |1\rangle)$, and an $n_0^{th}$-order correlation immune Boolean functions F.

2. Alice picks $\{a^{(i)} \in \{0,1\}^n | F(a^{(i)}) = b; i = 1, 2, \ldots, m\}$ and $\{\tilde{a}^{(i)} \in \{0,1\}^n | i = 1, 2, \ldots, m\}$ randomly and independently. Where $b$ is the value Alice committed.

3. Alice generates $\text{Blob}_4(b)$ and sends it to Bob.

  Open Phase:

4. Alice unveils $\{\tilde{a}^{(i)} \in \{0,1\}^n | i = 1, 2, \ldots, m\}$ to Bob.

5. Bob verifies that the blob he received is really coded with the bases $\{\tilde{a}^{(i)} \in \{0,1\}^n | i = 1, 2, \ldots, m\}$ through checking $F(a^{(i)})|_{i=1,2,\ldots,m} = b$ with his measurement result $\{a^{(i)} | i = 1, 2, \ldots, m\}$. If it is true, he accepts $b$ as the value Alice committed.

  Security Analysis: Binding Property. We first consider verifiability of $\text{Blob}_4(b)$. It can be seen that with a true set $\{\tilde{a}^{(i)} | i = 1, 2, \ldots, m\}$ Bob can get $\{a^{(i)} | i = 1, 2, \ldots, m\}$ with probability 1. He will find $F(a^{(i)})|_{i=1,2,\ldots,m} = b$. If Alice gives Bob a false set $\{\tilde{a}'^{(i)}|i=1,2,\ldots,m\}$ for changing her committed value from $b$ to $b \oplus 1$, the success probability of her cheat will be $\left(\frac{1}{2}\right)^m \to 0$ (for a sufficiently large $m$). Now we consider the inalterability of $\text{Blob}_4(b)$. What Alice can do are local operations to her probe qubits and classical communication for unveil her choices of coding bases. These cannot help her control the blob collapsing into a state corresponding to $\{\tilde{a}'^{(i)}|i=1,2,\ldots,m\}$ which satisfy $F(a'^{(i)})|_{i=1,2,\ldots,m} = b \oplus 1$ with a probability larger than $O(2^{-m})$, otherwise she can construct a scheme of superluminal communication. Suppose Alice can change Bob's measurement result form $\{a^{(i)}|F(a^{(i)})=b; i=1,\ldots,m\}$ to $\{a'^{(i)}|F(a'^{(i)})=b \oplus 1; i=1,\ldots,m\}$, they can do as follows:

1. One day Alice generates $\text{Blob}_4(b)$ by $\{\tilde{a}^{(i)}|i=1,2,\ldots,m\}$ and $\{a^{(i)}|F(a^{(i)})=b; i=1,\ldots,m\}$, then sends $b$, $\text{Blob}_4(b)$ and $\{\tilde{a}^{(i)}|i=1,2,\ldots,m\}$ to Bob together. They agree on communicating at 8 o'clock next morning.

2. Next morning at 8 o'clock, Alice and Bob execute their communication. If Alice wants to send 0, she does nothing before 8:00, and Bob will find the blob he received is really an encryption of $b$; If Alice wants to send 1, she changes Bob's blob before 8 o'clock and Bob will find the blob he has received is an encryption of $b' = b \oplus 1$.

Then we believe that Alice cannot control the changes of Bob's measurement result from $\{a^{(i)}|F(a^{(i)})=b; i=1,\ldots,m\}$ to $\{a'^{(i)}|F(a'^{(i)})=b \oplus 1; i=1,\ldots,m\}$. A variation of this problem is: Suppose F is a balanced Boolean function, then the probability of collapsing into a state fitting a given $b'$ is 0.5. Though Alice cannot control the random collapse to fit her object, she can understand the final state of the blob, and unveil a false basis for each wrong qubit. This strategy will increase her chance to success for each $n$-qubit string from 0.5 to 0.75, but cannot change the conclusion that her success probability of cheat approaches 0 exponentially.

Security Analysis: Concealing Property. It can be seen that without $\{\tilde{a}^{(i)}|i=1,2,\ldots,m\}$, Bob cannot get $\{a^{(i)}|F(a^{(i)})=b; i=1,\ldots,m\}$ correctly. His success probability for each $n$-qubit string is $\left(\frac{3}{4}\right)^n \to 0$. What should be mentioned is: though Bob cannot know any component of $a^{(i)}$ exactly in BB84-like protocol, he can get 75% components of each $a^{(i)}$ correctly via randomly chosen measurement bases. He can even get ~85% components of each $a^{(i)}$ via Breidbart attack. Consider of the sufficient large of parameter $m$, we should adopt correlation immune Boolean function in BB84-like protocol either.

Now let us see an alternative way to unconditionally secure bit commitment. Making use of the idea in B92-like protocol, we can construct a more practical bit commitment protocol via constructing first an unconditionally secure oblivious transfer protocol as follows:

1. Alice generates a qubit sequence according to her classical message $a = (a_1, a_2, \cdots, a_n)$ : $|\Psi_{a_1}\rangle, |\Psi_{a_2}\rangle, \ldots, |\Psi_{a_n}\rangle$ and sends it to Bob. Where $\{|\Psi_0\rangle, |\Psi_1\rangle\}$ the same as is described in (11).

2. Bob measures each qubits by a way which allows him to get a exact subset of $\{a_i | i = 1, 2, \cdots, n\}$.

It is obvious that Bob can only get a proper subset of $\{a_i | i = 1, 2, \cdots, n\}$ since his ability of differentiating between two nonorthogonal states is limited by $p_A = 1 - |\langle \Psi_0 | \Psi_1 \rangle|$. The probability of getting all components of a is $(p_A)^n \to 0$. Now let us see why Alice cannot know which subset is the one Bob get. Suppose Alice can get some information of Bob's subset, Alice and Bob can execute following protocol:

1. Alice sends $\{|\Psi_{a_i}\rangle | i = 1, 2, \cdots, 2n\}$ to Bob.

2. Bob chooses one of two choices as follows: one is measuring the first $n$ qubits with basis $\{|\Psi_0\rangle, |\bar{\Psi}_0\rangle\}$ and the second $n$ qubits with $\{|\Psi_1\rangle, |\bar{\Psi}_1\rangle\}$, where $|\bar{\Psi}_0\rangle$ and $|\bar{\Psi}_1\rangle$ is orthogonal states of $|\Psi_0\rangle$ and $|\Psi_1\rangle$ respectively; the other one is reverse.

It can be seen that Alice will know Bob's choice remotely if she can get any information about Bob's subset, then Alice and Bob realize a superluminal communication. We conclude that based on the principle of superposition of states and the law of causality we have proved the oblivious transfer protocol described above is unconditionally secure. Now let us construct a bit commitment protocol via this unconditionally secure oblivious transfer channel. Let F be an $n_0^{th}$-order correlation immune Boolean functions shared by the committer Alice and the receiver Bob.

Commit phase:

1. Alice picks randomly $\{a^{(i)} | F(a^{(i)}) = b;\ i = 1, \ldots, m\}$ and sends it to Bob via the secure oblivious transfer channel.

2. Bob gets proper subsets $\{s^{(i)} | i = 1, 2, \cdots, m\}$.

Open phase:

3. Alice unveils $\{a^{(i)} | i = 1, 2, \cdots, m\}$ to Bob via a classical channel.

4. Bob verifies that whether $s^{(i)}$ is really a subset of $a^{(i)}$ for each number $i$. If it is true, do the next step.

5. Bob calculates $F(a^{(i)})|_{i=1,2,\ldots,m}$. If $F(a^{(i)})|_{i=1,2,\ldots,m} = b$, he accepts $b$ as the value Alice committed.

It can be seen that one have to trace back to original B92-like bit commitment protocol if he wants to analyze thoroughly the security problem of this protocol, especially various possible cheat strategy. The advantages of this protocol are simple in concept and robust in practice, especially that it does not rely on Bob's ability of quantum storage.

*Discussions.* An unevadable question is: why Mayers-Lo-Chau no-go theorem does not work upon these protocols? Before answer this question, I would like to clarify first the relationship of two concepts: one concept is (such as) the trace distance between quantum states $\rho$ and $\sigma$ is infinitesimal; the other concept is the two states cannot be distinguished efficiently. It can be seen that these two concepts are not equivalent in some cases, such as in the case we described in B92-like scheme and BB84-like scheme. In those cases, because the security parameter $m$ and $n$ are both sufficiently large, any two blobs which have one qubit difference in each $n$-qubit string can be distinguished with probability $\to 1$, though the trace distance between these two blobs is

$$\begin{aligned} & D\left(\rho^B_{a'^{(1)}a'^{(2)}\cdots a'^{(m)}}, \rho^B_{a^{(1)}a^{(2)}\cdots a^{(m)}}\right)\Big|_{W_H(a'^{(i)} \oplus a^{(i)}) = 1} \\ & = \frac{1}{2}\frac{1}{mn}\text{tr}\left|\sum_{j=1}^{m}\sum_{i=1}^{n}\left(\left|\Psi_{a_i^{(j)}}\right\rangle\left\langle\Psi_{a_i^{(j)}}\right| - \left|\Psi_{a_i'^{(j)}}\right\rangle\left\langle\Psi_{a_i'^{(j)}}\right|\right)\right| \\ & \leq \frac{1}{2}\frac{1}{mn}\text{tr}\left(\sum_{j=1}^{m}\left|\sum_{i=1}^{n}\left(\left|\Psi_{a_i^{(j)}}\right\rangle\left\langle\Psi_{a_i^{(j)}}\right| - \left|\Psi_{a_i'^{(j)}}\right\rangle\left\langle\Psi_{a_i'^{(j)}}\right|\right)\right|\right) \\ & = \frac{1}{2n}\text{tr}\left\||\Psi_0\rangle\langle\Psi_0| - |\Psi_1\rangle\langle\Psi_1|\right\| \\ & = \frac{1}{2n}|(0,0,1) - (\sin 2A, 0, \cos 2A)| \\ & = \frac{1}{n}\sin A \sim O\left(\frac{1}{n}\right). \end{aligned} \qquad (16)$$

Now let us answer the unevadable question. It is clear that the protocols presented here belong $\rho_0^B \neq \rho_1^B$ case. The proof of no-go theorem was completed in this case by citing Uhlmann's theorem related with purification of mixed states. It can be seen that their proof can only leads to the conclusion that if the trace distance (for example) between states $\rho_0^B$ and $\rho_1^B$ is infinitesimal, the trace distance between the two states that Bob finally gets must be as well infinitesimal with the same or higher order. It is obvious that the example we just calculated is consist with this conclusion, though it is obvious as well that the two different concepts should not be confused.

In order to construct encryption function with verifiability as well as inalterability, we have to map one bit to $O(n^2)$ qubits. This leads to the distinction of any given pair of blobs having success probability approaches 1, then we get unconditionally secure binding property. It can be seen that any protocol of this character will not be a good concealing one except both $\rho_0^B$ and $\rho_1^B$ are multi-ford. We introduce correlation immune Boolean function to code committed bit, leads to mapping 1 bit into about $2^{O(n^2)}$ states, which makes Bob lost his object to compare with, even though he has the ability. This situation has not been considered in the proof of no-go theorem.

We can see that in our protocols Alice's cheats may sometimes lead to a first order infinitesimal, though the concealing is unconditionally secure. The reason is we have combined the non-orthogonal states coding with correlation immune Boolean transformation together. Concretely speaking, this result does not conflict with the formal proof of no-go theorem since the concealing we realized by quantum system is only $O\left(\frac{1}{n}\right)$, that is, we only require the non-orthogonal property of states guarantees that there is at least $n-n_0$ qubits that Bob cannot get for each $n$-qubit string. The improvement of concealing from $O\left(\frac{1}{n}\right)$ to $O(2^{-n})$ is obtained by means of correlation immunity of Boolean function, which guarantees that F($a$) is statistically independent to any $n_0$ components of $a$.

Briefly speaking, the proof of no-go theorem ignored the case in which the difference of two density matrices being infinitesimal does not always means one cannot distinguish them. This case may caused by the multi-ford property of $\rho_0^B$ and $\rho_1^B$ while we introduce a classical ensemble which mingled with the original quantum ensemble inseparably.

We conjecture that $O(n^2)$ qubits encryption function is essential for any non-interactive, unconditionally secure bit commitment protocols, and for any unconditionally secure, verifiable and unalterable strong quantum encryption function with $n$ bits input must have an output with $O(n^3)$ qubits.

It can be seen that Boolean function F can be substitute with function $(n, k)$: $\{0,1\}^n \to \{0,1\}^k$, where $k \sim O(1)$ as $n \to \infty$. This kind of substitution does not affect our conjecture.

It is easy to construct an unconditionally secure coin-flipping protocol on the top of these bit commitment protocols.

It should be mentioned that the loss of channel has no fatal affection to protocols presented here.

We would like to thank L. Hu and H. X. Xu for useful discussions. This work was supported by the National Natural Science Foundation of China (Grant No.60573051) and National Fundamental Research Program (Grant No.G2001CB309300).

____________________________________________

*Email address: yl@m165.com